\documentclass{elsart}
\usepackage{psfig}
\usepackage{natbib}
\def\home{}
\def\figdirprefix{\home}
\def\fig#1#2{\begin{figure}[htb]
\centerline{\psfig{figure=\figdirprefix#1.eps,width=8cm,angle=0,clip=}}
\caption{#2}
\label{#1}
\end{figure}}
\def\Fig#1#2{\fig{#1}{#2}Figure~\ref{#1}}
\begin{document}
\runauthor{Taylor}
%-------------------------------------------------------------------
\begin{frontmatter}
\title{Are Narrow-Line Seyfert 1s Really Strange?}
\author{Jason A. Taylor}
\address{taylor@taylorcapital.com \\ Taylor Capital, LLC \\ 3422 Oliver Street, NW \\ Washington, DC 20015 USA}
\begin{abstract}
Narrow-Line Seyfert 1s (NLS1s) are generally considered to be ``strange'' Active
Galactic Nuclei (AGNs). Surprisingly, this makes them very useful for constraining
models. I discuss what happens when one attempts to qualitatively fit the NLS1
phenomenon using the stellar wind model for AGN line emission (e.g., Kazanas
1989). The simplest way of narrowing profile bases of this model to the widths
observed in NLS1s is probably to lower the mass of the supermassive black hole.
In a flux-limited and redshift-limited data set, this is indeed similar to increasing
\( L/L_{\rm Edd} \). Because the broad line region (BLR) of the stellar line
emission model scales with the tidal radius of the stars, this model predicts
maximal BLR velocities of \( {\rm FWZI}\propto (L/L_{\rm Edd})^{-1/3} \). This
implies that the black holes of NLS1s are approximately \( 3^{3}=27 \) times
less massive than those in other Seyfert 1s if the stellar line emission model
is correct. Another consequence of increasing \( L/L_{\rm Edd} \) in this model
is that it results in an increase in the wind edge densities. NLS1 spectra appear
to support this result as well. Even the collateral features of NLS1s, such
as the line asymmetries and continuum properties, appear to be easily explained
within the context of this model. For better or worse, if the stellar wind line
emission is correct, NLS1s are not much stranger than other AGNs.

\end{abstract}
\end{frontmatter}
%-------------------------------------------------------------------------
NLS1s are generally considered to be ``strange'' sources. They are characterized
by low (\( \sim  \)500--2000 km s\( ^{-1} \)) H\( \beta  \) FWHMs, high (\( \sim  \)2.5-4.5)
0.1-2.4 keV photon indexes, scarcity of lobed radio emission, and low redward
H\( \beta  \) emission.

But NLS1s are AGNs nonetheless. We can, therefore, capitalize on the unusualness
of NLS1s to test the viabilities of AGN models. Incorrect models constructed by
theorists who were initially exposed only to ``normal'' AGN data  are likely
to fail once pushed outside the initial comfort zone of parameter space in which
theorists had lived. The correct model, on the other hand, should fit new and
unexpected data with ease merely by changing one or more existing parameters.
Ideally, the situation for the correct model/theorist would be comparable to,
e.g., a Landau and Lifshitz  expos$\acute{\rm e}$ of a situation in which a parameter could
be either real or imaginary, with the real values of the parameter corresponding
to the old data and the imaginary values corresponding to the new data.

For the NLS1 situation, we additionally know from Boroson \& Green (1992) that
the correct model must be able to fit both the ``normal'' AGN data and the NLS1
data upon the adjustment of only \emph{one} underlying parameter (or several
that are interdependent to the same effect). In my talk I showed what I found
when, as part of my thesis research (\S 4.14 in Taylor 1999), I tried to use
this result to qualitatively test the viability of just the stellar wind model.
Of course, NLS1s should permit powerful tests of \emph{all} the AGN BLR models,
and I had hoped to hear about analogous results for the other models (and other
BLR models in particular) at the conference. In this one respect I was disappointed;
I think only after such studies have been done will the full beauty of the NLS1
beast be brought to light.

In the stellar line emission model, the broad and narrow line emission is produced
in the winds of stars in the vicinity of the central black hole. The details of
this model are described elsewhere (Taylor 1999). In my talk I discussed the
model only in the context of NLS1s.  

The simplest way of lowering the profile widths to match NLS1s is probably
to lower the mass of the supermassive black hole. In a flux- and redshift-limited
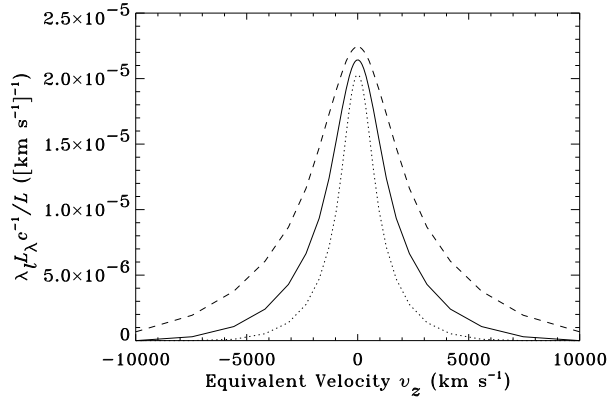
data set, this is indeed similar to increasing \( L/L_{\rm Edd} \). \Fig{jason_taylor_fig_1}{Theoretical, continuum-subtracted Ly\( \alpha  \)
line profiles for the stellar wind line emission BLR model. Solid line: model
2 (\( M_{\rm h}=3.0\times 10^{7}\, M_{\odot } \)). Dotted line: model 24 (\( M_{\rm h}=9.5\times 10^{6}\, M_{\odot } \)).
Dashed line: model 25 (\( M_{\rm h}=9.5\times 10^{7}\, M_{\odot } \)).} shows
the theoretical Ly\( \alpha  \) line profiles for three different models with
black hole masses that span one decade in parameter space. As one might naively
expect, the model with the smallest black hole (model 24) has the narrowest
profile. Since the luminosity was held constant in these models, model 24 is
also closest to the Eddington limit.

The results shown in Figure \ref{jason_taylor_fig_1} are not hard to obtain
analytically. In each of these three models, \( d\ln (An_{*})/d\ln r=-1.8<-1.0 \),
where \( An_{*} \) is the covering function. This means that the dominant contribution
towards their response covering functions occurs near \( r=r_{\rm t} \), where
\( r_{\rm t} \) is an average of the tidal radius of the relevant stars. Thus,
the delay of the summed line response function peak is comparable to the characteristic
delay of the line emission (defined as the integral of the delay-weighted response
function). In other words, because the covering functions are so steep in these
models, the covering is predominately a function of the inner radius (rather
than the outer radius). This simplifies the analysis dramatically. Let us make the crude approximation that the local line profile and distribution
function at the tidal radius are Gaussians proportional to \( {\rm exp}[-v^{2}/(2\sigma ^{2}_{\rm t})] \),
where \( v \) is the stellar velocity and \( \sigma _{\rm t} \) is the stellar
velocity dispersion at \( r=r_{\rm t} \). We then obtain for the velocity dispersions
of the broadest possible profile components (corresponding to emission from
\( r\simeq r_{\rm t} \)) 
\begin{equation}
\label{sigmaatrtidal}
\sigma _{\rm t}=\sqrt{\frac{GM_{\rm h}}{3r_{\rm t}}}=\sqrt{\frac{GM_{*}}{3R_{*}}}\left( \frac{M_{\rm h}}{\sqrt{2}M_{*}}\right) ^{1/3}\propto L^{1/3}\left( \frac{L}{L_{\rm Edd}}\right) ^{-1/3}.
\end{equation}

An analogous result can be obtained for the characteristic delays of the summed
line (defined as the sum of all BLR and NLR line emission) response function
peaks \( \tau _{\rm p\Sigma } \). This delay is also the minimum time scale
in the structure of the transfer function. \emph{If} the various nonlinear effects
are unimportant, these delays are 
\begin{equation}
\label{delayatrtidal}
\tau _{{\rm p}\Sigma }\simeq \frac{2R_{*}}{c}\left( \frac{2M_{\rm h}}{M_{*}}\right) ^{1/3}\propto L^{1/3}\left( \frac{L}{L_{\rm Edd}}\right) ^{-1/3}.
\end{equation}

Equations (\ref{sigmaatrtidal}) and (\ref{delayatrtidal}) are written in a
form to be used in understanding redshift- and flux-biased data sets. But note
that they imply
\begin{equation}
\label{tau-sigmarelation}
\tau _{\rm p\Sigma }=\frac{2^{7/6}R_{*}}{c}\left( \frac{GM_{*}}{3R_{*}}\right) ^{-1/2}\sigma _{\rm t}.
\end{equation}
 Equation (\ref{tau-sigmarelation}) is an important prediction of the stellar
wind AGN model. It states that the minimal structure size in the response function
of the summed line is proportional to the widths of the broadest components
of the lines. Though the proportionality constant between delay and velocity
depends upon the precise definitions assumed here and characteristics of the
stars in AGNs, the fundamental proportionality between delays and velocities
as they vary in AGNs with different black hole masses does not. These velocity-delay correlations are testable without making the questionable
assumption that few AGNs are sub-Eddington (cf. Boller, Brandt, \& Fink 1996).
It should be straightforward to falsify experimentally simply by plotting \emph{}\( \tau _{\rm p\Sigma } \)
and \emph{\( \sigma _{\rm t} \)} for several different AGNs to see whether
or not the relationship is linear. If the stellar BLR model is correct, these
relations imply that NLS1s are simply AGNs with relatively small supermassive
black holes. 

Moving on to the line-specific features of high \( L/L_{\rm Edd} \) models,
the continuum flux at \( r=r_{\rm t} \) that the clouds/winds are exposed to can
be written as \( F_{\rm c}\simeq L/(4\pi r_{\rm t}^{2})\propto U_{\rm t}n_{\rm Ht}\propto  \)
\( LM_{\rm h}^{-2/3}\propto L^{1/3}(L/L_{\rm Edd})^{2/3} \), where \( U_{\rm t} \)
is an average ionization parameter of the winds at the tidal radius and \( n_{\rm Ht} \)
is the Hydrogen density. Since thermal instabilities operating at the wind edges
should severely flatten the dependence of \( U \) upon \( F_{\rm c} \), the
edge density/pressure should get the lion's share of the \( F_{\rm c} \) functional
dependence. Thus, the densities of NLS1 BLR emission gas should be \( \sim 3^{2} \)
times higher than that of ``normal'' Seyferts. Spectral features of NLS1s,
such as the C {\small III}{]} line strengths, appear to be consistent with this. 

Other secondary features observed in NLS1s also appear to be in tune with what
one might expect in high-\( L/L_{\rm Edd} \) objects. For instance, unusually
high blueshifts of the high-ionization lines could be described reasonably well
if the simplifying assumption of spherically symmetric winds is eliminated and
the mean ionization parameter is slightly lower in NLS1s. And, at the risk of
stepping outside my own area of expertise, I speculate that the scarcity of
NLS1 jets might be attributable to the blocking off of the particle acceleration
mechanism by excessive high-\( z \) gas in thick high-\( L/L_{\rm Edd} \)
NLS1 disks. The possible high continuum variability amplitudes (discussed at
this meeting by, e.g., K. Leighly) seem to me to be expected in any positive-feedback
model of disk viscosity. 

In summary, while I agree that it might be difficult to make the whole NLS1
chapter look like something in a Landau and Lifshitz text, we at least have
some means of understanding NLS1s if we view them through the stellar line emission
model. And even if this model ends up being wrong, NLS1s still seem to me to
be comparable to the other AGN classes. I do not think they are strange at all.

\end{document}